\documentclass[conference]{IEEEtran}
\IEEEoverridecommandlockouts
\usepackage{cite}
\usepackage{algorithm,algorithmic} % for algorithm
\ifCLASSOPTIONcompsoc
\usepackage[caption=false, font=scriptsize, labelfont=sf, textfont=sf]{subfig}
\else
\usepackage[caption=false, font=footnotesize]{subfig}
\fi
\usepackage[pdftex]{graphicx}
\usepackage{balance}
\usepackage{embedfile}
\usepackage{caption}
\usepackage{epstopdf}
\usepackage{amsthm,amsmath,amssymb,amsfonts,bm}
\usepackage{algorithmic}
\usepackage{graphicx}
\usepackage{textcomp}
\usepackage{xcolor}
\usepackage{mathtools} % for \splitfrac

\newcommand{\source}{\mathsf{S}}
\newcommand{\des}{\mathsf{D}}
\newcommand{\PB}{\mathsf{PB}}

\newcommand{\R}{\mathsf{R}}
\newcommand{\PR}{\mathsf{PR}}
\newcommand{\PT}{\mathsf{PT}}

\newcommand{\w}{\mathbf{w}}

\newcommand{\varz}{\sigma^2}

\def\BibTeX{{\rm B\kern-.05em{\sc i\kern-.025em b}\kern-.08em
    T\kern-.1667em\lower.7ex\hbox{E}\kern-.125emX}}
\begin{document}

\title{\huge  An Efficient Deep CNN Design for EH Short-Packet Communications in Multihop Cognitive IoT Networks
}
	\author{
	\IEEEauthorblockN{
		Toan-Van Nguyen\IEEEauthorrefmark{1},
		Thien Huynh-The\IEEEauthorrefmark{2},
		Van-Dinh Nguyen\IEEEauthorrefmark{3},
		\\
		Daniel Benevides da Costa\IEEEauthorrefmark{4}\IEEEauthorrefmark{5},
		Rose Qingyang Hu\IEEEauthorrefmark{1}
		and~Beongku~An\IEEEauthorrefmark{6}}
	\IEEEauthorblockA{\IEEEauthorrefmark{1} Dept. of Electrical and Computer Engineering, Utah State University, Logan, UT, USA\\
		\IEEEauthorrefmark{2} ICT Convergence Research Center, Kumoh National Institute of Technology, Republic of Korea \\
		\IEEEauthorrefmark{3} SnT-Interdisciplinary Centre for Security, Reliability and Trust, University of Luxembourg, Luxembourg \\
		\IEEEauthorrefmark{4} AI \& Telecom Research Center, Technology Innovation Institute, 9639 Masdar City, Abu Dhabi, United Arab Emirates\\
		\IEEEauthorrefmark{5} Future Technology Research Center, National Yunlin University of Science and Technology (YunTech), Taiwan, R.O.C.\\
		\IEEEauthorrefmark{6} Dept. of Software and Communications Engineering, Hongik University, Republic of Korea}
	\IEEEauthorblockA{Emails: 
		\IEEEauthorrefmark{1}vannguyentoan@gmail.com, 
		\IEEEauthorrefmark{1}rose.hu@usu.edu,
		\IEEEauthorrefmark{2}dinh.nguyen@uni.lu,
		\IEEEauthorrefmark{3}thienht@kumoh.ac.kr,\\
		\IEEEauthorrefmark{4}\IEEEauthorrefmark{5}danielbcosta@ieee.org,
		\IEEEauthorrefmark{6}beongku@hongik.ac.kr}
}

\maketitle

\begin{abstract}
In this paper, we design an efficient deep convolutional neural network (CNN) to improve and predict the performance of energy harvesting (EH) short-packet communications in multi-hop cognitive Internet-of-Things (IoT) networks. Specifically, we propose a Sum-EH scheme that allows IoT nodes to harvest energy from either a power beacon or primary transmitters to improve not only packet transmissions but also energy harvesting capabilities. We then build a novel deep CNN framework with feature enhancement-collection blocks based on the proposed Sum-EH scheme to simultaneously estimate the block error rate (BLER) and throughput with high accuracy and low execution time. Simulation results show that the proposed CNN framework achieves almost exactly the BLER and throughput of Sum-EH one, while it considerably reduces computational complexity, suggesting a real-time setting for IoT systems under complex scenarios. Moreover, the designed CNN model achieves the root-mean-square-error (RMSE) of $\!{1.33\times10^{-2}}\!$ on the considered dataset, which exhibits the lowest RMSE compared to the deep neural network and state-of-the-art machine learning approaches.
\end{abstract}

%\vspace{-0.5em}
\vspace{-0.5em}
\section{Introduction}
Short-packet communication (SPC) has been specified as one of key technologies for the fifth generation (5G) networks because of its ability to provide high reliability and low end-to-end (e2e) latency \cite{parvez2018survey}. SPC supports a wide range of ultra-reliable low-latency communication (URLLC) applications, such as intelligent transportation systems, high-speed trains, drones, factory automation, and Internet-of-Things (IoT) networks \cite{ho2021short}. 
Recent works on URLLC mainly focused on SPCs with single-hop or dual-hop transmissions in factory automation, where the stringent requirements on reliability and latency are $99.999\%$ and $1$ ms, respectively \cite{ho2021short,saad2019vision}.

Cisco forecasted around 75 billion IoT devices in 2025, which poses a great challenge in allocating limited spectrum to those connected devices. By allowing sensor nodes to opportunistically access the vacant licensed spectrum, cognitive radio (CR) has been recognized as a spectrum-efficient networking solution for such a problem in dense IoT networks \cite{ren2018rf,ho2021short}. Besides deploying CR, radio-frequency energy harvesting (EH) technique is an attractive solution to supply stable energy sources in SPC-based systems \cite{makki2016wireless}, which provides long-term operation for the massive number of IoT devices in harsh environments and inaccessible locations. In \cite{lopez2017ultrareliable}, EH and data transmission processes were considered in ultra-reliable SPC scenarios. Recently, SPCs with EH have been investigated in a variety of network scenarios, e.g., ultra-reliable communications \cite{lopez2017ultrareliable}, wireless-powered IoT networks \cite{ho2021short}, and mission-critical IoT applications under finite blocklength regime \cite{wang2019secure}. However, SPCs in multi-hop cognitive IoT networks (MCINs) with multiple primary transceivers, which makes the performance analysis quite complex, have not been studied in these works. This prevents researchers from thoroughly examining the impact of finite blocklength and cognitive multi-hop transmissions under SPC study. 
%
%However, SPCs in multi-hop cognitive IoT networks (MCINs) assisted by deep learning have not been studied in these previous works, which prevents researchers from thoroughly evaluating the impact of finite blocklength regimes and multi-hop transmissions under SPC study. 
%
%This thus greatly restricts SPCs research for 5G and IoT networks under finite blocklength conditions with multi-hop transmissions, rendering a great challenge to the wireless communication community on how to accurately predict system performances under dynamic wireless environments.
%
%SPCs in multi-hop cognitive IoT networks (MCINs) assisted by deep learning have not been studied in these previous works.
%%
%%%

Recently, deep convolutional neural networks (CNNs) have been recognized as a powerful tool to solve various practical problems, such as resource allocation, queue management, and congestion control in modern wireless networks and IoT systems \cite{mao2018deep}. Since CNNs have the ability to accurately estimate high non-linear functions with low-complexity, it has been employed in a wide variety of interesting applications, such as relay selection, resource allocation, and channel estimations \cite{wang2019secure,wang2019decision}. CNN-based performance prediction helps expedite real-time settings in IoT networks since CNN models can precisely estimate desired performance metrics from high dimensional raw data even with dynamic environments and complex radio conditions, where a mathematical derivation is intractable. This paper carries out a fresh and first attempt to study SPCs with new energy harvesting strategies considering multiple primary transmitters ($\PT$s) and power beacon ($\PB$) in MCINs assisted by deep learning.
The main contributions of the paper can be summarized as follows:
\begin{itemize}
	\item It is proposed a Sum-EH scheme in multi-hop cognitive IoT networks, where the source and relay nodes can harvest energy from either $\PT$s or $\PB$ for their operations under short packet and cognitive radio constraints.
	\item To support real-time settings, we design an efficient CNN with feature enhancement-collection blocks to estimate system performance of MCINs, where the formulated block error rate (BLER) and throughput are converted into a multi-output regression problem for a low-latency and high accuracy estimation.
	\item It is shown through simulation results that CNN framework achieves almost the same BLER and throughput with that of Sum-EH one, while it drastically reduces complexity and execution time. The Sum-EH scheme also outperforms other EH strategies, such as Max-EH and primary transmitter-based-EH schemes, which corroborates the efficiency of the proposed EH method.
	\item It is also revealed that the designed CNN model provides the lowest root-mean-square-error (RMSE) compared to the deep neural network (DNN) and state-of-the-art machine learning (ML) approaches, arising as an excellent estimator for performance prediction.
\end{itemize}

\textit{Mathematical Notations}: Boldface represents vector and $\|.\|$ symbolizes the Frobenius norm. $\mathbb{E}\{\cdot\}$ and $(\cdot)^H$ are the expectation operator and transpose conjugate, respectively.
%%
%\vspace{-0.5em}
\section{System Model} \label{sec_system}
\subsection{System Description and Operation}
\begin{figure}[!th]
	\centering
	\includegraphics[width=0.9\linewidth]{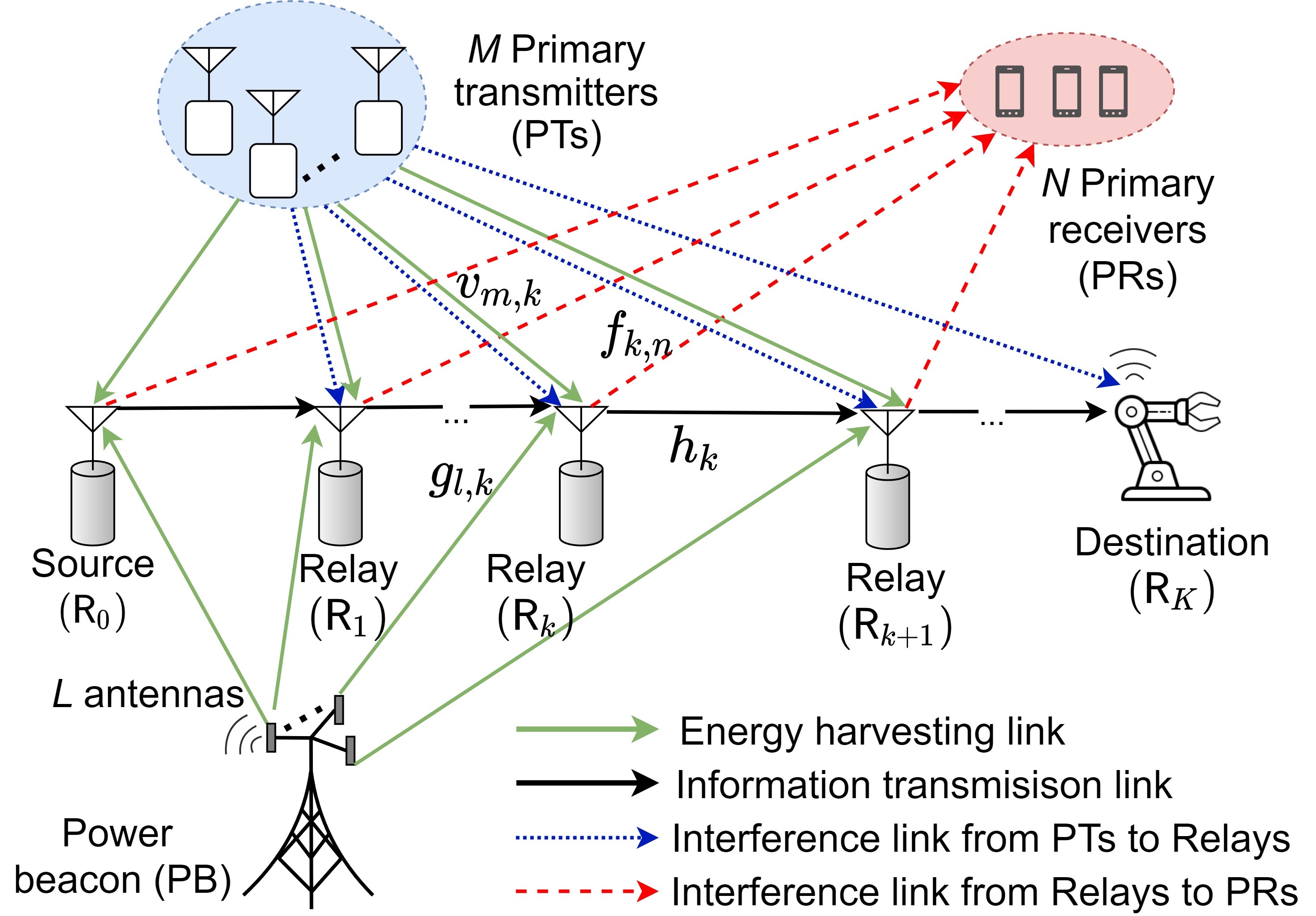}
	\caption{The proposed SPC in MCINs with EH capabilities.}
	\label{fig1}
\end{figure}
We consider a secondary multi-hop cognitive IoT network cognitively operating with a primary network, as shown in Fig.~\ref{fig1}. A single-antenna source ($\R_0$) sends its short-packets to a single-antenna destination ($\R_K$) through $K$-hop transmissions, where each relay $\R_k$, with $k=1,\ldots,K-1$, uses decode-and-forward protocol to transmit its received signal to the next hop in the presence of $N$ primary receivers ($\PR$s). Since source and relays can be seen as energy-limited IoT devices, they must harvest energy from a $\PB$ with $L$ antennas and from $M$ $\PT$s for their operations. We assume that perfect channel state information is available at each terminal.

We denote by $g_{l,k}$, $v_{m,k}$, $h_{k}$, and $f_{k,n}$ the channel coefficients from $l$-th antenna at $\PB$ to $\R_{k}$, with ${l=1,\ldots,L}$, from $m$-th $\PT$ to $\R_{k}$, with ${m=1,\ldots,M}$, from $\R_{k-1}$ to $\R_{k}$, and from $\R_k$ to $n$-th $\PR$, with ${n=1,\ldots,N}$, respectively. The propagation channels experience both large-scale path loss and small-scale fading. Thus, the channel coefficient is modeled as $u=\sqrt{G}\tilde{u}$, with $\tilde{u} \in \{ g_{l,k}, v_{m,k}, h_{k}, f_{k,n} \}$, where $G$ and $\tilde{u}$ present the large-scale path loss and small-scale fading, respectively.
%which is generated as independent circularly symmetric complex Gaussian random variable with distribution $\mathcal{CN}(0,1)$, where complex circularly symmetric Gaussian distributions. 
The small-scale channel is modeled by Rayleigh fading such that the channel gain $|u|^2$ follows an exponential distribution with parameter $\lambda_u$. On the other hand, the large-scale path loss is modeled as ${G= {\sigma^{PL}}{{(d/d_0)}^{-\tt PL}}}$, where $d$, $\tt{PL}$, $d_0$, and $\sigma^{PL}$ denote the distance between two nodes, the path loss exponent, the reference distance, and the power attenuation at $d_0$, respectively.

We consider SPCs in multi-hop cognitive IoT networks with two consecutive phases, which include EH and information transmission (IT) phases.
We denote by $m$ and $T$ the total number of channel uses and duration of each channel use, respectively; thus, the $K$-hop transmissions in one block time are denoted by $m T$. Moreover, the number of channel uses of EH is denoted by $n_E$; thus, the period of $n_E T$ is spent for EH and the remaining one, $(m-n_E)T$, is used for IT process.
%%
%\vspace{-0.5em}
\subsection{Energy Harvesting Phase}
Considering the deployment of $\PB$ and $\PT$s, three energy harvesting strategies can be presented as follows:
%
%\vspace{-0.4em}
\subsubsection{Primary Transmitter-based Energy Harvesting Scheme}
In this scheme, the relay nodes harvest energy only from the $\PT$s since the $\PB$ is located very far from the multi-hop network.
The energy harvested by $\R_{k-1}$, with $\!k=1,\ldots,K$, from $M$ $\PT$s can be expressed as
\begin{align}
\label{Eq:Ek}
{E}_{k-1} =  {\eta n_E T} P_{\PT} \sum_{m=1}^{M} |v_{m,k-1}|^2,
\end{align}
where $\eta \in(0,1)$ and $P_{\PT}$ are the energy conversion efficiency and the transmit power of each $\PT$, respectively.
The IT process is performed after the EH process during the time period of $t_D=(m-n_E)T/K$, where the transmit power of $\R_{k}$ can be calculated from \eqref{Eq:Ek} as
\begin{align}
\label{Eq:PkE:PT}
{P}_{k-1}^{\rm PT} = \kappa P_{\PT} \sum_{m=1}^{M} |v_{m,k-1}|^2,
\end{align}
where $\kappa  \triangleq {K\eta n_E}/{(m-n_E)}$ and $P_{\PT}$ presents the transmit power of each $\PT$ in the primary network.
\subsubsection{Max-Energy Harvesting Scheme}
In this scheme, relay nodes harvest the highest amount of energy between $\PT$s and $\PB$ for their operation. The EH of $\R_{k-1}$ can be expressed as
\begin{align}
\label{Eq:Ek:Max}
\!{E}_{k-1\!} \!=\!  \!{\eta n_E T}\!\max \!\Big\{ \!P_{\PB} |\mathbf{g}_{k-1}^H \! \w_{k-1\!}|^2\!, P_{\PT}\! \sum_{m=1}^{M} |v_{m,k-1\!}|^2 \!\Big\},
\end{align}
where $P_{\PB}$ denotes the transmit power of each antenna at $\PB$, $\w_{k-1}$ and $\mathbf{g}_{k-1}$ are the $L\!\times 1$ beamforming and $L\!\times 1$ channel coefficient vectors between $\PB$ and $\R_{k-1}$, respectively. $\PB$ employs an efficient maximal-ratio transmission criterion based on beamforming design for powering the source and relay nodes, where it is formulated as
$\w_{k-1\!} \!=\! {\mathbf{g}_{k-1\!}}/{\|\mathbf{g}_{k-1\!}\|}$ \cite{wang2019secure}. Thus, the transmit power of $\R_{k-1}$ can be expressed as
\begin{align}
\label{Eq:PkE:Max}
{P}_{k-1}^{\rm Max} = \kappa \max\Big\{P_{\PB} \|\mathbf{g}_{k-1}\|^2, P_{\PT} \sum_{m=1}^{M} |v_{m,k-1}|^2 \Big\}.
\end{align}
\subsubsection{Sum-Energy Harvesting Scheme}
The Sum-EH criterion is deployed to combine the harvested energy from either $\PT$s or $\PB$ for the operations of source and relay nodes, where the EH at $\R_{k-1}$ can be expressed as
\begin{align}
\label{Eq:Ek:Sum}
{E}_{k-1} =  {\eta n_E T}\Big( P_{\PB} \|\mathbf{g}_{k-1}\|^2 + P_{\PT} \sum_{m=1}^{M} |v_{m,k-1}|^2 \Big).
\end{align}
Based on \eqref{Eq:Ek:Sum}, the transmit power of $\R_{k-1}$ can be expressed as
\begin{align}
\label{Eq:PkE:Sum}
{P}_{k-1}^{\rm Sum} = \kappa  \Big(P_{\PB} \|\mathbf{g}_{k-1}\|^2 + P_{\PT}\sum_{m=1}^{M} |v_{m,k-1}|^2 \Big).
\end{align}
%
%\vspace{-0.5em}
\subsection{Information Transmission Phase}
To guarantee the quality-of-service of primary communications, the transmit power of relays in multi-hop networks should be lower than a predefined threshold $I_{th}$ required by the PRs. The transmit power of $\R_{k-1}$ can be expressed as
\begin{align}
P_{k-1}^{\rm Sch} = \min\Big\{ {P}_{k-1}^{\rm Sch},{I_{th}}/{\underset{n=1,\ldots,N}{\max} |f_{k-1,n}|^2} \Big\},
\end{align}
where $\rm Sch \in \{ PT, Max, Sum\}$.

Upon receiving the signal transmitted from $R_{k-1}$, the instantaneous signal-to-noise ratio (SNR) at $\R_{k}$ can be formulated as
\begin{align}\label{gm:Sum}
\gamma_k^{\rm Sch} = \frac{P_{k-1}^{\rm Sch}}{\sigma^2}\frac{|h_k|^2}{P_{\PT} \sum_{m=1}^{M}|v_{m,k}|^2},
\end{align}
where $P_{\PT}  \sum_{m=1}^{M}|v_{m,k}|^2$ indicates the interference from $M$ PTs at $\R_{k}$.
%
%\vspace{-0.5em}
\section{Performance Evaluation} \label{secPerformance}
Considering finite blocklength transmission condition, the performance of multi-hop network is analyzed and evaluated in terms of average block error rate (BLER) and throughput.
%
%\vspace{-0.5em}
\subsection{Block Error Rate}
The source transmits a packet of $b$ information bits (message size) via $K$ hops over the blocklength ${n_D=(m-n_E)/K}$, with $\!{n_D > 100}\!$ \cite{makki2016wireless}, and the e2e SNR equals to $\gamma_{e2e}$, with transmission rate given by $r = b/n_D$. Thus, the average BLER at hop $k$ can be tightly approximated by \cite{lopez2017ultrareliable,wang2019secure}
\begin{align}
\label{Eq:EP}
\varepsilon_{k}^{\rm Sch} \approx  \mathbb{E}\bigg\{ Q\bigg( \frac{C(\gamma_{k}^{\rm Sch} ) - r}
{\sqrt{V(\gamma_{k})/n_D} } \bigg) \bigg\} ,
\end{align} % \mathbb{E}
where $Q(.)$ represents the Gaussian Q-function, $C(x) \triangleq \log_2(1+x)$ and $V(x) \triangleq \Big(1-\frac{1}{(1+x)^2}\Big)(\log_2e)^2$ denote the Shannon capacity and the channel dispersion, respectively.
The e2e BLER of multi-hop network can be expressed as
\begin{align}
\label{PoutBRS}
\varepsilon_{e2e} = 1-\prod_{k=1}^{K}\big[1-\varepsilon_k^{\rm Sch}  \big].
\end{align}
\vspace{-0.6em}
\subsection{Throughput}
We consider the delay-limited transmission scenario, where the throughput is determined by evaluating the e2e BLER.
For a fixed data transmission rate $R_{th}$ bits per channel use (BPCU) and the effective communication time ${(m-n_E)T/K}$ over total transmission time $m T$, the e2e throughput in delay-limited transmission mode can be expressed as
\begin{align} \label{TP:limited}
\tau_{e2e} \!=\! R_{th} {(m-n_E)}(1- \varepsilon)/{mK}.
\end{align}
To support real-time settings, we design an efficient CNN model for BLER and
throughput predictions in the next section.
%\vspace{-0.5em}
\section{Deep CNN Design for Performance Prediction}
%
%\vspace{-0.5em}
\subsection{Description of CNN}
We design a deep CNN to accurately estimate the throughput in the considered system setup.
The network is initialized by an input layer with size $1 \times 15$ to process input data, which includes the number of antennas at $\PB$ ($L$), the number of relays ($K$), the number of PTs ($M$), the number of PRs ($N$), the positions of PTs ($x_\PT,y_\PT$), PRs ($x_\PR,y_\PR$), and $\PB$ ($x_\PB,y_\PB$), the transmit powers of $\PB$ ($P_{\PB}$), $\PT$ ($P_{\PT}$), the threshold at the PRs ($I_{th}$), the channel uses for EH ($n_E$), and the target data rate ($R_{th}$). Based on the range of input variables shown in Table \ref{tab1}, the dataset is generated over $140,000$ samples for training and testing.
%We have numerically observed that this number of samples is sufficient to obtain high accurate predictions in most cases.

\begin{table}[!th]
	\centering
	\caption{Input variables and their values for CNN.}
	\label{tab1}
	\setlength{\tabcolsep}{7pt}
	{\renewcommand{\arraystretch}{1.2}
		\begin{tabular}{|c|c||c|c||c|c|}
			\hline
			\bf Variable &  \bf Value  & \bf Variable &  \bf Value  & \bf Variable &  Value \\\hline
			$L$ 	& $[1,6]$ & $x_\PT$ & $[18,20]$ & $y_\PB$ & $[8,10]$  \\
			$K$ 	& $[1,6]$ & $y_\PT$ & $[28,30]$ & $P_{\PB}$ & $[0,30]$ \\
			$M$ 	& $[1,6]$ & $x_\PR$ & $[18,20]$ & $I_{th}$ & $[0,30]$ \\
			$N$ 	& $[1,6]$ & $y_\PR$ & $[18,20]$ & $P_{\PT}$ & $[0,40]$ \\
			$n_E$ 	& $[1,6]$ & $x_\PB$ & $[8,10]$  & $R_{th}$ & $[1,2]$ \\\hline
	\end{tabular}}
	%\footnote{van text}
\end{table}

\begin{figure}[!th]
	\centering
	\includegraphics[width=0.75\linewidth]{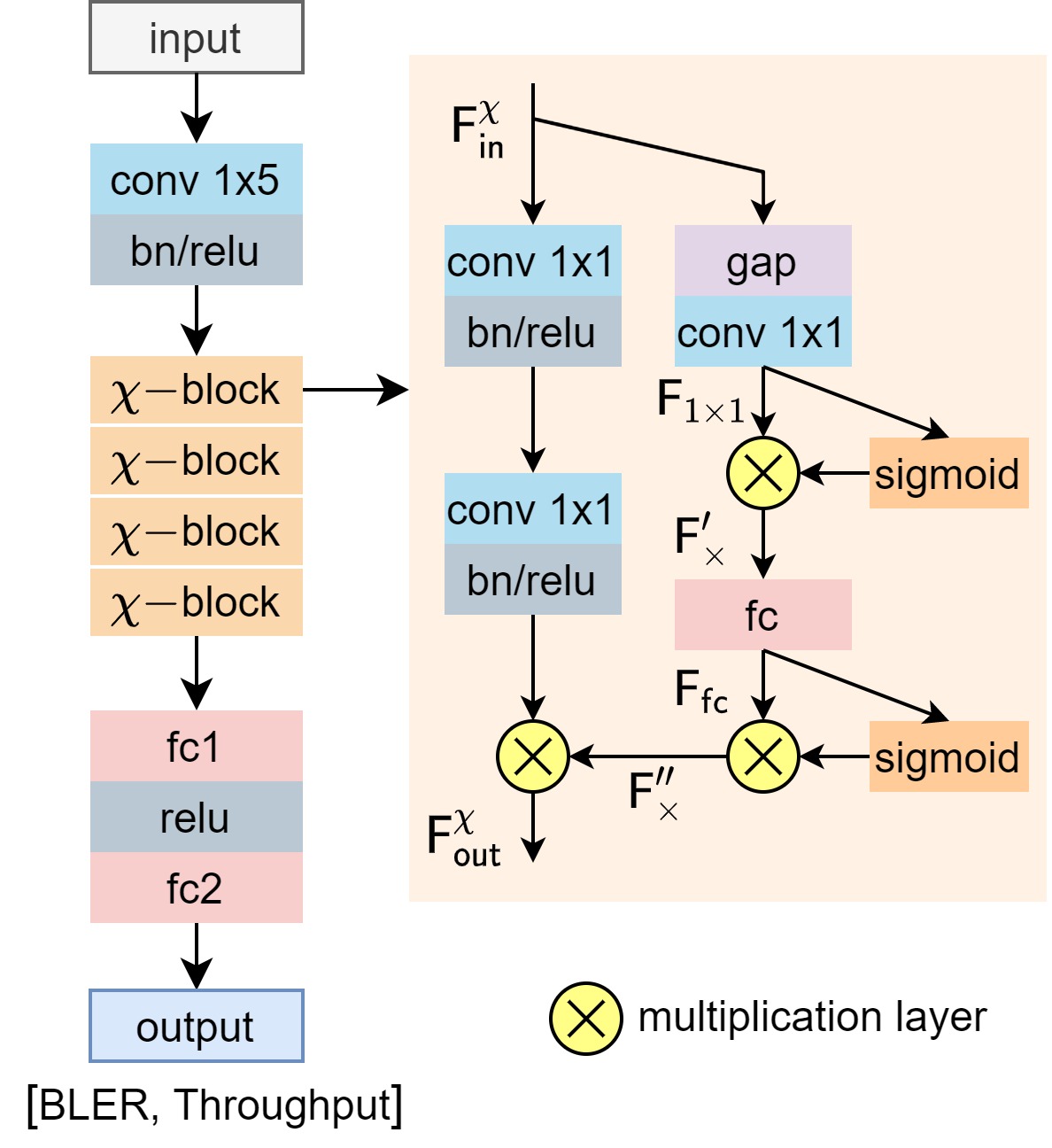}
	\caption{Deep CNN for BLER and throughput estimation with overall architecture and the detailed structure of blocks.}
	\label{fig2}
\end{figure}

With regard to the network architecture, there are four feature enhancement-collection blocks, denoted by $\chi-\sf block$, which are sequentially linked together to extract the intrinsic features as the correlations between system variables, as illustrated in Fig.~\ref{fig2}.
Our deep network starts with the input size of $1\times15$ indicating the system variables in Table \ref{tab1}. We next specify a convolutional ($\sf conv$) layer with kernel size of $1\times 5$, followed by a batch normalization ($\sf bn$) layer and a rectified linear unit ($\sf ReLu$) layer to effectively enrich the feature diversity.
To emphasize the high impact features resulted by the preceding layers, $\chi-\sf block$ is designed to possibly improve accuracy of prediction. 
As the core of network architecture, each $\chi-\sf block$ has three multiplication layers. 

%Notably, each \textit{conv} layer in blocks is followed by a batch normalization layer and a rectified linear unit (ReLU) activation layer.
Given the input of $\chi-\sf block$ $\mathsf{F}_{\sf in}^{\chi}$, its operating principle can be described as follows.
At the first multiplication layer, the output is obtained as
\begin{equation}
% \mathsf{F}_{\times}'   = \mathcal{G}_{1\times 1}\big(\mathsf{F}_{\sf in}^{\chi}\big) \odot \mathcal{S}\big( \mathcal{G}_{1\times 1}\big(\mathsf{F}_{\sf in}^{\chi}\big)\big),
\mathsf{F}_{\times}'   = \mathsf{F}_{1\times1} \odot \mathcal{S}\big( \mathsf{F}_{1\times1}\big),
\end{equation}
where $\mathsf{F}_{1\times1} = \mathcal{G}_{1\times 1} \big( \mathcal{P}_{\sf gap} \big(\mathsf{F}_{\sf in}^{\chi}\big) \big)$, in which $\mathcal{P}_{\sf gap}$ and $\mathcal{G}$, respectively, indicate the global average pooling and the $1 \times 1$ convolution operation; $\mathcal{S}$ and $\odot$ denote the sigmoid function and element-wise multiplication, respectively. The output $\mathsf{F}_{\sf x}'$ is passed through a fully connected ($\sf fc$) layer, where the result $\mathsf{F}_{\sf fc}$ is used as input of the second multiplication layer, i.e.,
\begin{equation}
\mathsf{F}_{\times}''   = \mathsf{F}_{\sf fc} \odot \mathcal{S}\big( \mathsf{F}_{\sf fc} \big).
\end{equation}
The output of $\chi-\sf block$ is identical to the output of the third multiplication layer, which can be expressed as
\begin{equation}
		\mathsf{F}_{\sf out}^{\chi}  = \mathsf{F}_{\sf x}'' \odot \mathcal{G}_{1\times 1}\Big(\mathcal{G}_{1\times 1}\Big(\mathsf{F}_{\sf in}^{\chi} \Big)\Big) .
\end{equation}
%
%In $\sf relu$ layers, the nonlinearity is performed for every input element $z$ by the following equation
%\begin{align} \label{mtA}
%\sf{ReLU}(z) = \left\{ {\begin{array}{*{10}{c}}
%	\begin{array}{cr}
%	z,\\
%	0,
%	\end{array}&\begin{array}{l}
%	\textrm{if}~  z>0, \\
%	\textrm{if}~ z<0.
%	\end{array}
%	\end{array}} \right.
%\end{align}
The $\sf{ReLu}$ function greatly accelerates the convergence of optimization algorithm during training process compared to the sigmoid or tanh activation functions \cite{krizhevsky2012imagenet}.
%To save the network computation when going deeper with more \textit{conv} layers, we arrange a max pooling (\textit{maxpool}) layer with the pool size of $1\times2$ and the stride of $\left( 1,2\right)$ in block-B and block-C to reduce the spatial size of feature maps.
%Remarkably, we deploy an effective structure to collect the high-level features summarized at multi-scale representations.
%In details, besides connecting blocks in cascade, the output of each block is also passed to a global average pooling (\textit{gap}) layer and a fully connected (\textit{fc}) layer.
%Subsequently, the outputs from these \textit{fc} layers are then combined together via a general \textit{concat} layer as selective feature collection.
The network is finalized with a $\sf relu$ layer between two $\sf fc$ layers, denoted by $\sf fc1$ and $\sf fc2$ in Fig.~\ref{fig2}, wherein the number of neurons in $\sf fc2$ is identical to the number of estimation variables at the output. Here, the BLER and throughput variables need to be estimated and they are arranged into a vector $\textbf{y}\triangleq [\varepsilon_{e2e} \,\, \tau_{e2e}]$.
With regard to the layer configurations, we specify 64 kernels in each $\sf conv$ layer and 64 neurons in each $\sf fc$ layer. Based on the universal approximation theory \cite{hornik1989multilayer}, the heuristic-based scaling approach is applied in this design, where the number of kernels and neurons are heuristically chosen by experimental trials until the minimal training error is achieved \cite{tan2020efficientdet}.
It is worth noting that the zero padding is automatically added when processing convolution with different kernel sizes to keep the spatial size of output feature maps unchangeable.

For the regression problem, the loss function indicating the error between predicted and expected values, which can be expressed as \cite{zhang2020neural}
\begin{align}
\label{loss}
\mathcal{L} = \frac{1}{M} \sum_{m=1}^{M}( \textbf{y}_{(m)} - \tilde{\textbf{y}}_{(m)})^2,
\end{align}
where $M$ is the number of training samples, and $\textbf{y}_{(m)}$ and $\tilde{\textbf{y}}_{(m)}$ are the expected and predicted values, respectively.
The weights and biases are iteratively updated during the backpropagation procedure by using the adaptive moment (Adam) estimation optimization algorithm to minimize the loss function of the entire training set.
%\vspace{-0.5em}
\subsection{Real-Time Prediction}
Once the offline training is completed, the resulting CNN model consisting of weights and biases can be represented in a compact mapping function as $\mathcal{F}(.)$. In general, when the CNN is well trained, it can provide real-time and highly accurate predictions. We use the resulting CNN model to predict the throughput value whenever any new information is available at the input. In particular, we input serially each sample, which is arranged as a vector $\mathbf{x} \triangleq [L,K,M,N,x_\PT,y_\PT,x_\PR,y_\PR,\\x_\PB,y_\PB,P_\PB,I_{th},P_\PT,n_E,R_{th}]$, and the resulting CNN will output the predicted BLER and throughput sorted into vector $\tilde{\textbf{y}}\triangleq [\tilde\varepsilon_{e2e} \,\, \tilde\tau_{e2e}]$, which can be expressed as 
\begin{align} \label{mapping}
\tilde{\textbf{y}} =  \mathcal{F}(\mathbf{x}).
\end{align}
Through a low-latency inference process in \eqref{mapping}, the BLER and throughput can be predicted by the CNN model within a short time. Since the capacity of CNN can be improved by going a deeper (more hidden layers) or wider (more neurons/hidden units) network, the settings of CNN can be aptly designed to achieve the lowest error during training process.
If the predicted throughput is not close to the expected one, the CNN will need to be re-trained with new appropriate settings until achieving the smallest error in \eqref{loss}.
%\vspace{-0.7em}
\section{Simulation Results and CNN Predictions} \label{secNumercial}
In this section, we present illustrative examples to evaluate the BLER and throughput of the proposed system model. Monte-Carlo simulations and CNN prediction results are provided to validate our designed approach. We consider a two-dimensional plane, where $\source$, $\des_l$, $\R_k$, and $\PB$ are located at coordinates $(0,0)$, $(25,0)$, $(k/K,0)$, and $(10,10)$, respectively. We set the reference distance ${d_0=1}$ m and the pathloss at $d_0$ $\sigma^{PL}=-30$ dB. Unless otherwise stated, we set in simulations the number of antennas at $\PB$ as $L = 4$, the number of relays as $K = 4$, the number of PTs as $M = 4$ and PRs as $N=3$, the energy conversion efficiency as $\eta = 0.8$, the pathloss exponent as $\beta = 2.6$, the total channel uses as $m=1500$ with $256$ bits, the channel uses for EH as $n_E=500$, and the normalized noise variance as $\varz = 1$. To prevent the overfitting issue while network goes deeper, $5$ hidden layers and $120$ hidden neurons are selected for DNN model \cite{ho2021short}.
The entire dataset is split into $80\%$ for training and the remainder for testing. The hold-out cross-validation method is used, where the model is trained on the training set and evaluated on the test set.
We also heuristically set $70$ epochs for training of the CNN model to prevent the overfitting on the training set and achieve a good accuracy on the test set. The weights and biases are randomly initialized using Adam optimizer with the gradient decay factor of $0.95$. The initial learning rate is set as $10^{-3}$ (dropped $90\%$ after $20$ epochs). The performance is measured on a system equipped by a 3.60-GHz CPU, 32GB RAM, and a single NVIDIA GeForce GTX 1060 6GB GPU. 
%The prediction error of CNN is reported on the test set. 
%
%\vspace{-0.5em}
\subsection{RMSE Evaluation}
To evaluate the estimation performance of the proposed CNN-based estimation scheme for predicting the target BLER and throughput, the RMSE is used to measure the differences between actual values and predicted ones over the entire test set. The RMSE can be calculated by using the samples in the test set as follows:
\begin{align}
\label{RMSE}
\mathrm{RMSE} = \sqrt{\mathbb{E}\{(\textbf{y} - \tilde{\textbf{y}})^2\}}.
\end{align}
%
%\noindent where $\mathbb{E}\{.\}$ denotes the expectation.
We compare the RMSE of the CNN-based model with that of ML-based regression models, such as SVM \cite{jian2016budget}, decision tree learning \cite{wang2019decision}, and ensemble learning \cite{ren2016ensemble}. It is noted that such ML-based regression models share the same dataset with the proposed CNN one.
We follow the standard $\mathsf{k}$-fold cross-validation ($\mathsf{k}=5$) to measure the performance of traditional ML regression models \cite{jian2016budget}.
%
%To demonstrate the effectiveness of CNN-based model in DNS scheme
\begin{figure}[!th]
	\centering
	\includegraphics[width=0.9\linewidth]{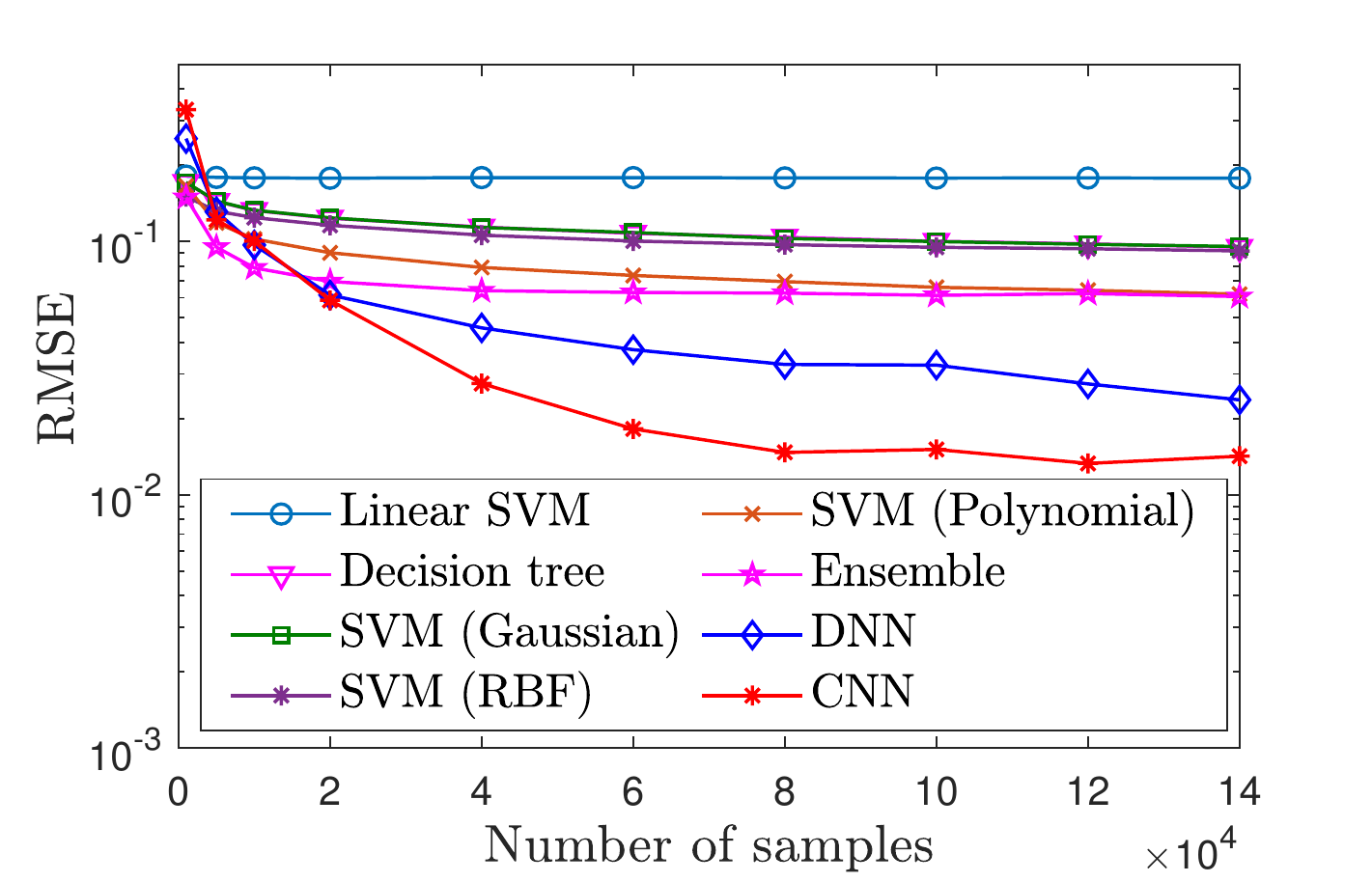}
	\caption{RMSE versus the number of samples.}
	\label{Fig_ML_and_CNN}
\end{figure}

We first present the RMSE versus the number of samples with different regression models, as shown in Fig. \ref{Fig_ML_and_CNN}. The linear SVM regression model has the highest RMSE, while the proposed CNN model gets the lowest RMSE, showing the best performer. The reason is that the linear SVM model is infeasible to estimate a moderate-to-high-dimensional dataset, resulting in the lowest performance. The SVM model with Polynomial, Gaussian, and FBF kernels yield almost a similar RMSE result since they share the same backbone architecture.
The RMSE values of the ensemble, DNN, and the proposed CNN models are progressively reduced and eventually reach $0.0606$, $0.0237$, and $0.0142$, respectively, over the entire test set. 
The proposed CNN model has the ability to map the original dataset into a higher dimensional space through the versatile design of $\chi-\sf block$ and attention connection. This shows the beneficial feature of deep learning approach for effectively dealing with big communication datasets. On the contrary, the RMSE of the SVM, decision tree, and ensemble models is nearly unchanged even when the number of samples is increased, showing that they have poor predictive ability in large datasets.

%\vspace{-0.2em}
\begin{figure}[!th] 
	\centering
	\includegraphics[width=0.9\linewidth]{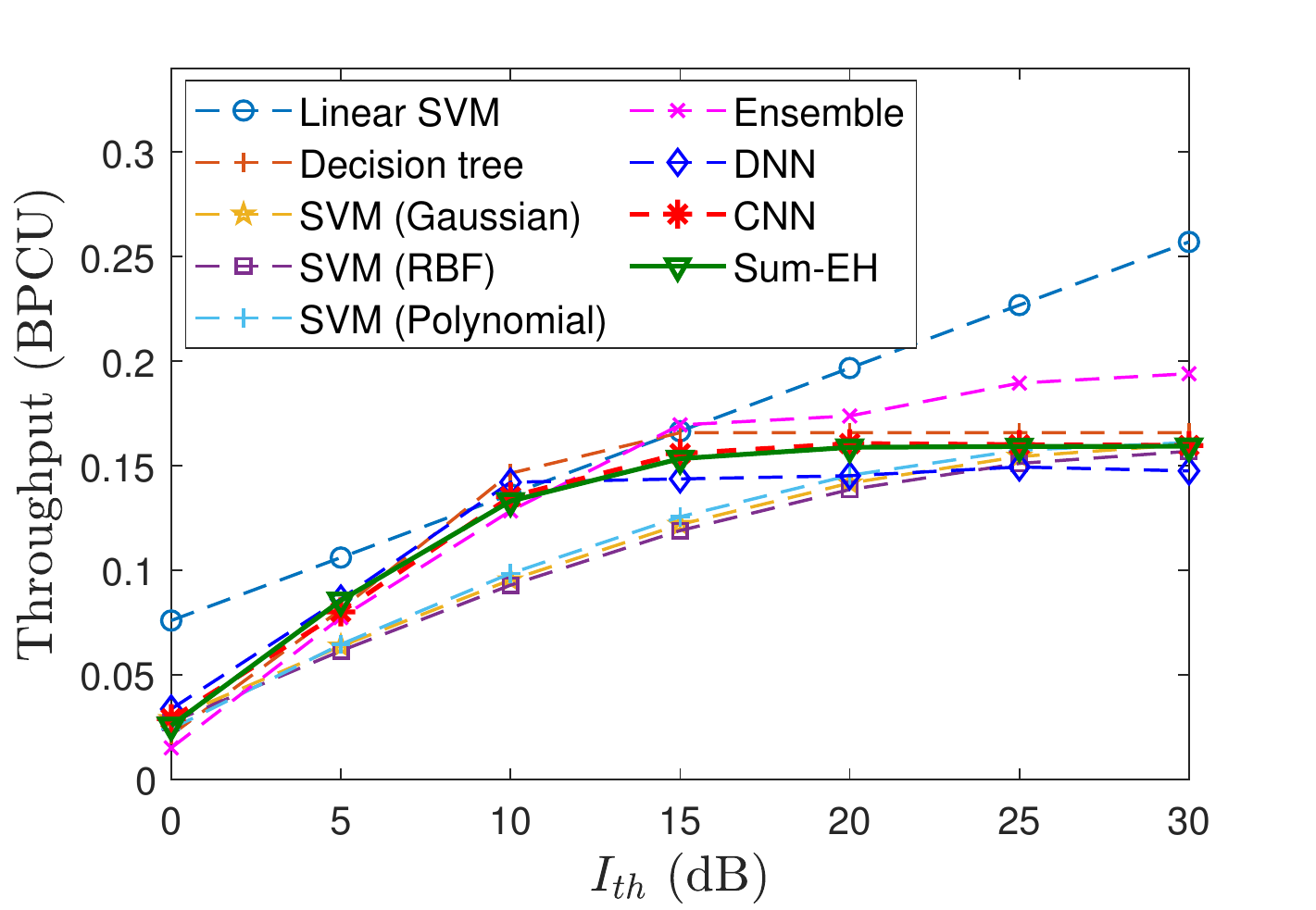}
	\caption{Throughput prediction results of the proposed CNN and different ML-based estimation schemes.}
	\label{Fig_TP_ML_and_CNN}
\end{figure}
The effects of different ML-based estimation schemes on the ability of throughput prediction are shown in Fig. \ref{Fig_TP_ML_and_CNN}.
It is observed that the proposed CNN-based estimation framework provides the best fit curve of the throughput to the Sum-EH scheme while the linear SVM-based estimation one has a high error prediction and fails to evaluate the throughput. The decision tree, ensemble, and DNN-based estimation schemes perform better than SVM (linear, RBF, and polynomial kernels) one; however, they cannot exactly predict throughput at high SNRs since their nature is shallow learning network. Contrarily, a novel $\!\chi-\sf\! {block}$ architecture designed for deep CNN allows it to explore and learn the inter-feature correlations, which prevents the network from vanishing gradient by deploying batch normalization layer. Owing to this fact, the CNN-based estimation framework demonstrates the best prediction ability among DNN and ML-based estimation schemes.
\vspace{-0.6em}
\subsection{BLER and Throughput Evaluation}
\begin{figure}[!th] 
	\centering
	\includegraphics[width=0.9\linewidth]{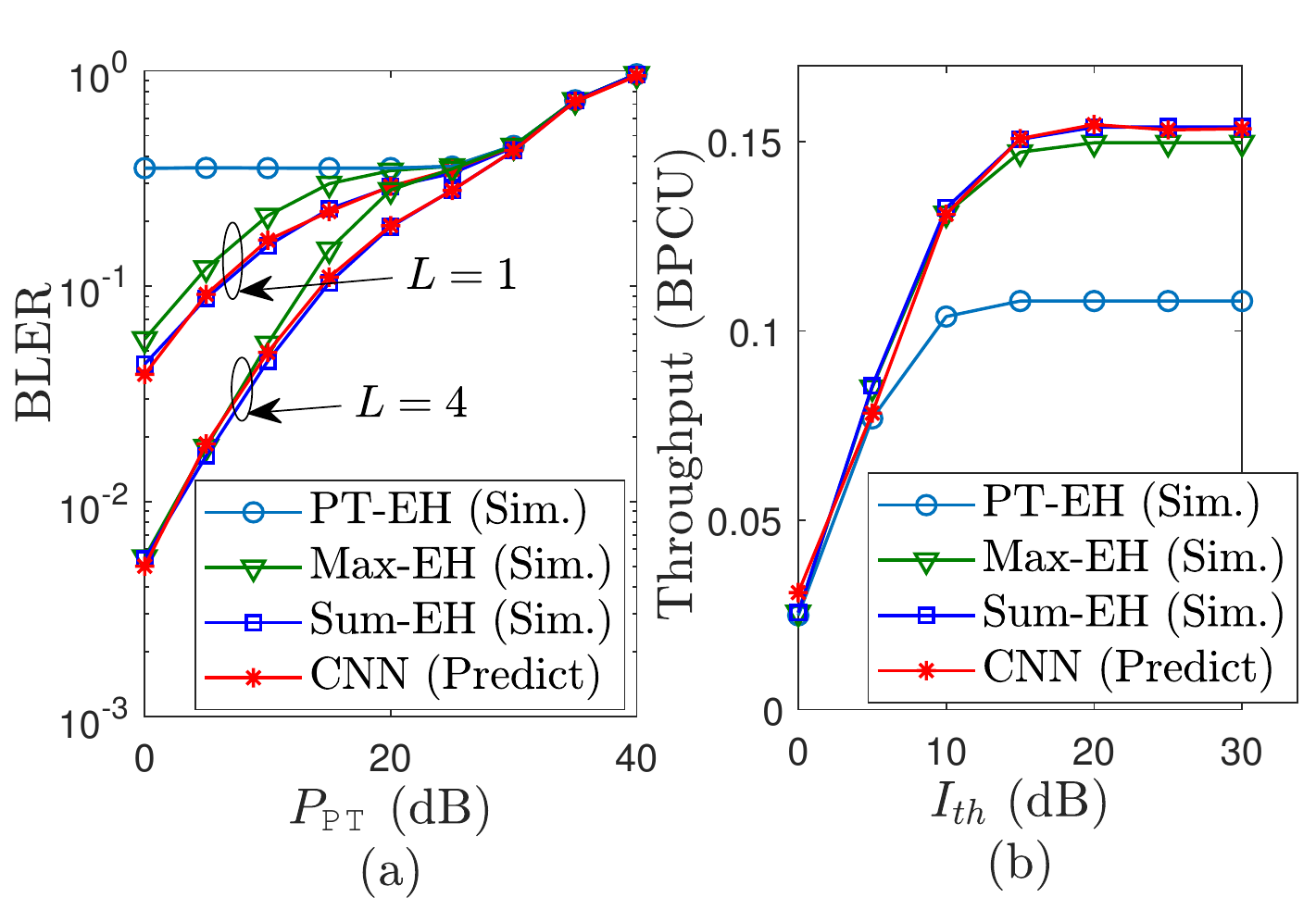}
	\caption{Average BLER and throughput of different schemes.}
	\label{BLER_TP}
\end{figure}
In Fig. \ref{BLER_TP}(a), we show the BLER of all schemes with different values of $L$. When the number of antennas at the $\PB$ increases, the EH capability enhances at relay nodes as \eqref{Eq:PkE:Max} and \eqref{Eq:PkE:Sum}, thus improving the BLER for Sum-EH and Max-EH schemes. However, when $P_{\PT}$ increases, its harmful effect counterbalances the benefits gained from EH; thus, the BLER of all schemes suffers outage at $P_{\PT}=40$ dB.
It is observed in Fig. \ref{BLER_TP}(b) that the throughput of all schemes increases when $I_{th}$ is large. The source and relays will have more energy budget if the allowable interference level $I_{th}$ at the PRs is large, leading to the enhancement of system throughput. Moreover, the Sum-EH scheme gives the best performance while the PT-EH scheme is the lowest performer. The reason is that relay nodes in Sum-EH scheme can harvest a total of energy from PTs and PB while the PT-EH scheme mainly harvests from PTs. Moreover, the negative effects caused by the interference from PTs on relays offsets the positive effects given by EH from PTs; thus, PT-EH has poor BLER and throughput performances.
It is also evident that the Monte-Carlo simulation results of the Sum-EH scheme perfectly coincide with the CNN prediction ones, verifying our excellent deep model design.
\vspace{-0.5em}
\subsection{Reliability and Latency Evaluation}
The reliability and latency~\cite{lopez2017ultrareliable,ho2021short} can be defined as 
\begin{align}
&{\rm Reliability} = (1-\varepsilon_{e2e})\times 100\% ,\\
%\end{align} 
%The latency can be defined as \cite{ho2021short}
%\begin{align}
&{\rm Latency} =  {(m-n_E)T}/{(1-\varepsilon_{e2e})},
\end{align}
where $T=3 \mu s$ \cite{lopez2017ultrareliable}.
\begin{figure} [!th] \vspace{-0.2em}
	\centering
	\includegraphics[width=0.9\linewidth]{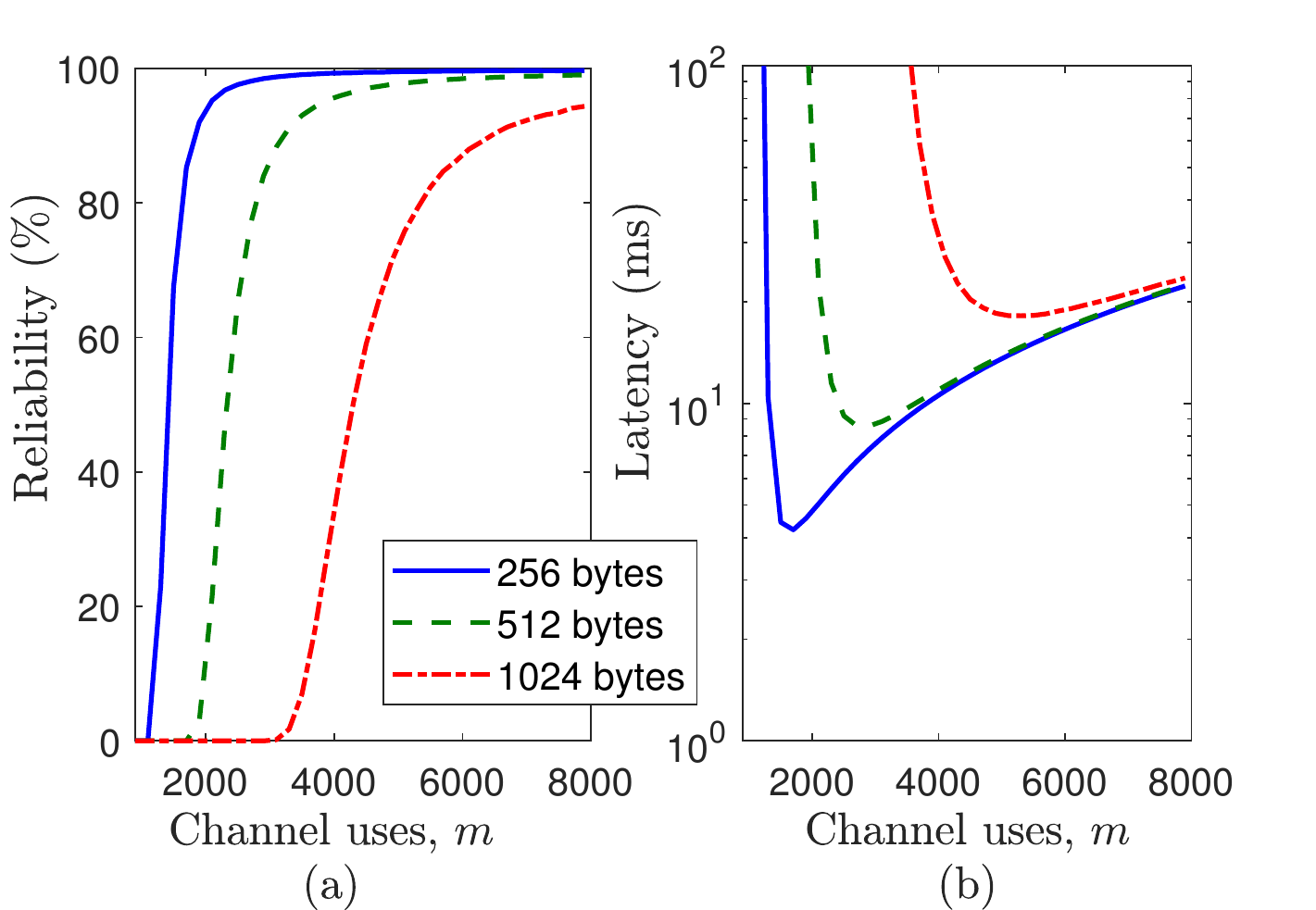}
	\caption{The reliability and latency of Sum-EH scheme.}
	\label{Fig_Latency_Reliability} %\vspace{-0.2em}
\end{figure}

In Fig.~\ref{Fig_Latency_Reliability}, the message with 256 bytes has higher reliability and lower latency than long one (i.e., with 512 or 1024 bytes). To achieve high reliability (over $99.99\%$), the 512-byte message can be framed into packets by using $6000$ channel uses, as shown in Fig. \ref{Fig_Latency_Reliability}(a), but its latency is about $18$s, as shown in Fig.~\ref{Fig_Latency_Reliability}(b), which exceeds the maximum latency budget for URLLC applications \cite{saad2019vision}. The 1024-byte message is transmitted with the latency of $20$s and reliability of $90\%$ at $6000$ channel uses, which does not meet the stringent requirements of latency and reliability. Thus, the long-messages are not able to support URLLCs in IoT networks and low-latency transmission in factory automation.
\vspace{-0.3em}
\subsection{Execution Time Evaluation}
%\vspace{-0.3em}
\begin{table}[!th]
	\centering
	\caption{Execution time comparison for the throughput evaluation among Monte-Carlo simulation (Sim.), and CNN method, with $I_{th}=20$ dB, $P_{\PT}=12$ dB, and $P_{\PB}=10$ dB.}
	\label{tab2}
	\setlength{\tabcolsep}{4pt}
	{\renewcommand{\arraystretch}{1.1}
		\begin{tabular}{|c|c|c|c|c|}
			\hline
			Scenarios $\{K,L,M,N\}$ &	Sim. & CNN & RMSE \\	\hline
			$\{2,3,4,3\}$ &  $13.266$s  & $0.34474$s & $0.0013$ \\ \hline
			$\{3,4,5,4\}$ &  $24.918$s  & $0.96858$s & $0.0025$ \\ \hline
			$\{4,5,6,5\}$ &  $39$s      & $0.91551$s & $0.0037$ \\ \hline
	\end{tabular}}
	%\footnote{van text}
\end{table}
Finally, we evaluate the execution time of the throughput prediction in Table \ref{tab2}. Each sample of Monte-Carlo is obtained by averaging $5\times 10^5$ independent channel realizations. When the network scale increases in terms of the number of devices $K,M,N$ and antennas $L$, the CNN still guarantees an execution time of less than $1$s in all scenarios, while the execution time of simulation method increases with the network scale and it consumes $39$s for the last scenario.
These results reveal the excellent ability of the proposed CNN framework in dealing with large scale network settings. 
%Thus, the CNN prediction contributes to expediting real-time configuration in practical multi-hop cognitive IoT networks.
\vspace{-0.3em}
\section{Conclusions} \label{secConclusions}
In this paper, we proposed a novel deep CNN-based relay selection scheme in multi-hop cognitive IoT networks to evaluate and improve the e2e BLER and throughput. Simulation results showed that the proposed CNN-based estimation scheme achieved almost exactly the throughput of Sum-EH one, which by its turn outperformed the Max-EH, and PT-EH schemes. Furthermore, the design of deep CNN model shown the perfect estimation capability with the smallest RMSE compared to DNN and machine learning approaches. In future works, we will study the hybrid relay-reflecting intelligent surfaces for multi-hop IoT systems embedding various advanced deep learning models to solve the joint power allocation, beamforming, and relay selection problem in future IoT cognitive wireless networks.
\vspace{-0.3em}
\section*{Acknowledgment}
\vspace{-0.2em}
The work of B. An was supported by National Research Foundation of Korea (NRF) grant funded by the Korea government (MSIT) (NRF-2019R1A2C1083996). The work of V.-D. Nguyen was supported in part by the ERC AGNOSTIC project, ref. H2020/ERC2020POC/957570.
Prof. Beongku An is the corresponding author.
%\section*{References}
%\balance
\def\baselinestretch{0.95}
\bibliographystyle{IEEEtran}
\bibliography{References_GC20}

\begin{thebibliography}{10}
\providecommand{\url}[1]{#1}
\csname url@samestyle\endcsname
\providecommand{\newblock}{\relax}
\providecommand{\bibinfo}[2]{#2}
\providecommand{\BIBentrySTDinterwordspacing}{\spaceskip=0pt\relax}
\providecommand{\BIBentryALTinterwordstretchfactor}{4}
\providecommand{\BIBentryALTinterwordspacing}{\spaceskip=\fontdimen2\font plus
\BIBentryALTinterwordstretchfactor\fontdimen3\font minus
  \fontdimen4\font\relax}
\providecommand{\BIBforeignlanguage}[2]{{%
\expandafter\ifx\csname l@#1\endcsname\relax
\typeout{** WARNING: IEEEtran.bst: No hyphenation pattern has been}%
\typeout{** loaded for the language `#1'. Using the pattern for}%
\typeout{** the default language instead.}%
\else
\language=\csname l@#1\endcsname
\fi
#2}}
\providecommand{\BIBdecl}{\relax}
\BIBdecl

\bibitem{parvez2018survey}
I.~Parvez, A.~Rahmati, I.~Guvenc, A.~I. Sarwat, and H.~Dai, ``A survey on low
  latency towards {5G: RAN}, core network and caching solutions,'' \emph{IEEE
  Commun. Surveys Tuts.}, vol.~20, no.~4, pp. 3098--3130, 4th Quart. 2018.

\bibitem{ho2021short}
C.~D. Ho, T.-V. Nguyen, T.~Huynh-The, T.-T. Nguyen, D.~B. da~Costa, and B.~An,
  ``Short-packet communications in wireless-powered cognitive {IoT} networks:
  Performance analysis and deep learning evaluation,'' \emph{IEEE Trans. Veh.
  Technol.}, vol.~70, no.~3, pp. 2894--2899, 2021.

\bibitem{saad2019vision}
W.~Saad, M.~Bennis, and M.~Chen, ``A vision of {6G} wireless systems:
  Applications, trends, technologies, and open research problems,'' \emph{IEEE
  Network}, vol.~34, no.~3, pp. 134--142, May 2019.

\bibitem{ren2018rf}
J.~Ren, J.~Hu, D.~Zhang, H.~Guo, Y.~Zhang, and X.~Shen, ``{RF} energy
  harvesting and transfer in cognitive radio sensor networks: Opportunities and
  challenges,'' \emph{IEEE Commun. Mag.}, vol.~56, no.~1, pp. 104--110, 2018.

\bibitem{makki2016wireless}
B.~Makki, T.~Svensson, and M.~Zorzi, ``Wireless energy and information
  transmission using feedback: Infinite and finite block-length analysis,''
  \emph{IEEE Trans. Commun.}, vol.~64, no.~12, pp. 5304--5318, Dec. 2016.

\bibitem{lopez2017ultrareliable}
O.~L.~A. L{\'o}pez, H.~Alves, R.~D. Souza, and E.~M.~G. Fern{\'a}ndez,
  ``Ultrareliable short-packet communications with wireless energy transfer,''
  \emph{IEEE Signal Process. Lett.}, vol.~24, no.~4, pp. 387--391, Apr. 2017.

\bibitem{wang2019secure}
H.-M. Wang, Q.~Yang, Z.~Ding, and H.~V. Poor, ``Secure short-packet
  communications for mission-critical {IoT} applications,'' \emph{IEEE Trans.
  Wireless Commun.}, vol.~18, no.~5, pp. 2565--2578, May 2019.

\bibitem{mao2018deep}
Q.~Mao, F.~Hu, and Q.~Hao, ``Deep learning for intelligent wireless networks: A
  comprehensive survey,'' \emph{IEEE Commun. Surv. Tutor.}, vol.~20, no.~4, pp.
  2595--2621, 4th Quart. 2018.

\bibitem{wang2019decision}
X.~Wang, ``Decision-tree-based relay selection in dualhop wireless
  communications,'' \emph{IEEE Trans. Veh. Technol.}, vol.~68, no.~6, pp.
  6212--6216, Jun. 2019.

\bibitem{krizhevsky2012imagenet}
A.~Krizhevsky, I.~Sutskever, and G.~E. Hinton, ``Imagenet classification with
  deep convolutional neural networks,'' in \emph{Adv. Neural Inf. Process Syst.
  25}, Dec. 2012, pp. 1097--1105.

\bibitem{hornik1989multilayer}
K.~Hornik, M.~Stinchcombe, and H.~White, ``Multilayer feedforward networks are
  universal approximators,'' \emph{Neural Netw.}, vol.~2, no.~5, pp. 359--366,
  1989.

\bibitem{tan2020efficientdet}
M.~Tan, R.~Pang, and Q.~V. Le, ``Efficientdet: Scalable and efficient object
  detection,'' in \emph{Proc. of IEEE/CVF Conf. Computer Vision and Pattern
  Recognition (CVPR)}, Seattle, USA 2020, pp. 10\,781--10\,790.

\bibitem{zhang2020neural}
Z.~Zhang, Y.~Lu, Y.~Huang, and P.~Zhang, ``Neural network-based relay selection
  in two-way {SWIPT}-enabled cognitive radio networks,'' \emph{IEEE Trans. Veh.
  Technol.}, vol.~69, no.~6, pp. 6264--6274, Jun. 2020.

\bibitem{jian2016budget}
L.~Jian, S.~Shen, J.~Li, X.~Liang, and L.~Li, ``Budget online learning
  algorithm for least squares {SVM},'' \emph{IEEE T. Neur. Net. Lear.},
  vol.~28, no.~9, pp. 2076--2087, Sept. 2016.

\bibitem{ren2016ensemble}
Y.~Ren, L.~Zhang, and P.~N. Suganthan, ``Ensemble classification and
  regression-recent developments, applications and future directions,''
  \emph{IEEE Comput. Intell. M.}, vol.~11, no.~1, pp. 41--53, 2016.

\end{thebibliography}
%\vspace{12pt}
\end{document}